\documentclass[12pt, pdftex]{iopart}

\expandafter\let\csname equation*\endcsname\relax

\expandafter\let\csname endequation*\endcsname\relax
\usepackage{amssymb,amsthm}
\usepackage{amsmath}
\usepackage{graphicx}
\usepackage{color}
\usepackage{ulem}
\usepackage{url}
\usepackage[pagewise]{lineno}
\usepackage{arydshln}
\usepackage[hang,small,bf]{caption}
\usepackage[subrefformat=parens]{subcaption}
\usepackage{multirow}
\usepackage{here}
\usepackage{bbm}

\DeclareRobustCommand{\erase}{\bgroup\markoverwith{\textcolor{red}{\rule[.5ex]{2pt}{0.4pt}}}\ULon}

\newcommand{\two}{I\hspace{-1.2pt}I}

\begin{document}

\newif\ifdraft
\draftfalse

\title[]{Extension of the characterization method for non-Gaussianity in gravitational wave detector with statistical hypothesis test}

%
%






\author{Shunsei Yamamura$^{1, 2}$, Hirotaka Yuzurihara$^2$, Takahiro Yamamoto$^2$, Takashi Uchiyama$^2$}
\address{$^1$Department of Physics, University of Tokyo, 7-3-1 Hongo, Bunkyo-ku, Tokyo 113-0033, Japan}
\address{$^2$Institute for Cosmic Ray Research (ICRR), KAGRA Observatory, The University of Tokyo, Kamioka-cho, Hida City, Gifu 506-1205, Japan}
\ead{yamamura@icrr.u-tokyo.ac.jp}


\begin{abstract}%
In gravitational wave astronomy, non-Gaussian noise, such as scattered light noise disturbs stable interferometer operation, limiting the interferometer's sensitivity, and reducing the reliability of the analyses.
In scattered light noise, the non-Gaussian noise dominates the sensitivity in a low frequency range of less than a few hundred Hz, which is sensitive to gravitational waves from compact binary coalescence.
This non-Gaussian noise prevents reliable parameter estimation, since several analysis methods are optimized only for Gaussian noise.
Therefore, identifying data contaminated by non-Gaussian noise is important.
In this work, we extended the conventional method to evaluate non-Gaussian noise, Rayleigh statistic, by using a statistical hypothesis test to determine a threshold for non-Gaussian noise.
First, we estimated the distribution of the Rayleigh statistic against Gaussian noise, called the background distribution, and validated that our extension serves as the hypothetical test.
Moreover, we investigated the detection efficiency by assuming two non-Gaussian noise models.
For example, for the model with strong scattered light noise, the true positive rate was always above 0.7 when the significance level was 0.05.
The results showed that our extension can contribute to an initial detection of non-Gaussian noise and lead to further investigation of the origin of the non-Gaussian noise.
\end{abstract}


\maketitle


\section{Introduction}

In 2015, the first gravitational wave (GW) event called GW150914 was detected by the Laser Interferometer Gravitational-wave Observatory (LIGO) Hanford and Livingston detectors during the first observing run (O1) \cite{PhysRevLett.116.061102,Abbott_2016}.
During the second observing run (O2), the GW170817 event was observed by LIGO \cite{aLIGO} and Virgo in Italy \cite{AdVirgo}, which was the first report of GW signals from coalescing binary neutron stars \cite{GW170817, Multimessenger}.
So far, three observing runs called O1, O2, and O3 have been conducted, and 90 GW events have been reported in the GW catalog \cite{GWTC3-arxiv}.
The fourth observing run (O4) by LIGO, Virgo, and KAGRA \cite{KAGRA1} started in May 2023.

Strain data is the GW detector's output in which it is possible to include the GW. 
Such data contain much noise in the detector such as seismic noise, thermal noise, laser radiation pressure noise, and shot noise.
These noise sources limit the sensitivity of the detector in different frequencies.
For example, in the case of estimated sensitivity limits of KAGRA, the radiation pressure noise mainly affects the frequency range below a hundred Hz, while the shot noise affects frequencies higher than 300 Hz.
Therefore, Fourier transforms are used to evaluate the data in the frequency domain.

The noise can be classified into two categories, Gaussian noise and non-Gaussian noise.
In Gaussian noise, the real and imaginary parts of the Fourier coefficients follow the Gaussian distribution independently.
By contrast, the Fourier coefficients of non-Gaussian noise deviate from the Gaussian distribution.
Examples of Gaussian noise include laser radiation pressure and shot noise, while examples of non-Gaussian noise include line noise, glitches, and scattered light noise.
Line noise is distributed at narrow band frequencies, and is caused by suspension resonances, main power lines, injected calibration lines, etc.
Glitch is a non-stationary noise that lasts within a few seconds.
Scattered light noise occurs when a small fraction of the main beam gets scattered, hits a moving surface, and recouples with the main beam. 
This noise dominates the sensitivity in the low frequency range of less than a few hundred Hz \cite{Impact_of_upconverted_scattered_light}, where signals from the coalescence of binary black holes can be found \cite{GW190521}.

Several  analysis methods to detect for GWs are optimized for Gaussian noise \cite{PE_for_CB,pyCBC,pycbc_inference_PE}.
Therefore, the results of analyses can be improved by evaluating the Gaussianity, which represents how close the data is to Gaussian noise, and identifying non-Gaussian noise \cite{yamaT, sensitive_test_non_Gaussianty}.
To characterize the non-Gaussian behavior in strain data, several approaches have been developed and investigated \cite{Virgo_detchar, LIGO_detchar_O3, Yuzurihara:detchar, PTEP_detchar}.

The Rayleigh statistic \cite{gwpy, gwpy_doi} has been used in LIGO to evaluate the Gaussianity.
Virgo has also proposed and used a method to evaluate the Gaussianity, called rayleighSpectro \cite{Virgo_detchar}.
We used the Rayleigh statistic in this paper.
The Rayleigh statistic is sensitive to various non-Gaussian noise sources, because it is independent of the noise model.
Due to low computational costs, it is often used for online monitoring of strain data.
While the Rayleigh statistic evaluates the Gaussianity quantitatively, setting the threshold for determining the presence of non-Gaussian noise is difficult.

Therefore, we extended the Rayleigh statistic by introducing a statistical hypothesis test.
A hypothesis test can statistically infer whether a null hypothesis is true.
By setting the null hypothesis that the noise is Gaussian noise, the presence of non-Gaussian noise can be judged from the result of the Rayleigh statistic.
As a result, the extension can serve as a means to determine the threshold according to the false positive rate.

This extension provides two advantages.
First, it is independent of a noise model when searching for non-Gaussian noise, although many previous studies rely on a noise model to judge the data quality \cite{Virgo_detchar,LIGO_detchar_O3,DQSEGDB}.
Therefore, the extension can provide a new indicator for an initial detection of non-Gaussian noise, even for the noise that are not modeled. 
The initial detection can then facilitate detailed investigation—for example, a coincidence study with auxiliary channels and checking the power in the time-frequency region using a Q transform \cite{qtransform}.

Secondly, the detection efficiency of the Rayleigh statistic against a specific noise model can be evaluated using the receiver operating characteristic (ROC) curve.
In this paper, we simulated two non-Gaussian noise models. 
One is an abstract model using transient Gaussian noise, and the other is a concrete model of scattered light detected before.
The detection efficiency was evaluated against these simulated noises while changing some parameters.

Section \ref{sec:Method} introduces a statistical hypothesis test, the Rayleigh statistic, and the extension of the Rayleigh statistic. 
Section \ref{sec:evaluation} describes details of non-Gaussian noise models.
Here, we assume non-Gaussian models and evaluate the ROC curve using non-Gaussian data.

\section{Method and Validation}
\label{sec:Method}

In this section, we introduce a statistical hypothesis test, the definition of the Rayleigh statistic, 
and a method to extend the Rayleigh statistic by using the hypothesis test.
The results of this extension are compared with the theoretical expectation against Gaussian noise and validated.

\subsection{Hypothesis test}
\label{sec:hypothesis test}
A hypothesis test is a method of statistical inference judging whether the data follows a hypothesis.
The first step is to set the null hypothesis, $H_0$, the alternative hypothesis, $H_1$, the significance level, $\alpha$, and the test statistic, $R$.  
The distribution of $R$ when $H_0$ is true (hereafter referred to as the background distribution) is required to calculate a $p$-value.
The background distribution can be numerically estimated by the Monte Carlo method.
In the two-tailed test that tests whether a test static is out of a confidence interval of values,
a $p$-value is computed by comparing  $R$ and the samples of the background distribution defined as
\begin{align}
p = 
\begin{cases}
\displaystyle
     \frac{2}{n_\mathrm{bg}}\sum_{i=1}^{n_\mathrm{bg}} \mathbbm{1}_{\{R_i \geq R\}} 
     \hspace{20pt}(R \geq R_m)\\
\displaystyle
    \frac{2}{n_\mathrm{bg}}\sum_{i=1}^{n_\mathrm{bg}} \mathbbm{1}_{\{R_i < R\}}
    \hspace{20pt}(R < R_m),
\end{cases}
\label{eq:p-value}
\end{align}
where $R_i$ is the $i$-th value of the sorted samples of background distribution, $R_m$ is the sample median of the background distribution, $n_\mathrm{bg}$ is the number of samples in the background distribution, and 
$\mathbbm{1}_{\{\cdot\}}$ is a function with value 1 when $\{\cdot\}$ is true and 0 when $\{\cdot\}$ is false.

If the $p$-value is less than $\alpha$, the case is called significant; 
$H_0$ is rejected and $H_1$ is accepted. 
If the $p$-value is higher than $\alpha$ and $H_0$ is not rejected, it does not mean $H_0$ is true, and the decision is reserved.

Table \ref{table: confusion matrix} summarizes four categories classified based on the the results of the hypothesis test.
The column indicates whether $H_0$ is true, 
and the row indicates whether $H_0$ is rejected in the test.
The classifications are as follows.
When $H_0$ is true, it is classified as true negative (TN) if $H_0$ is not rejected in the test and false positive (FP) if $H_0$ is rejected. 
When $H_0$ is false, it is classified as false negative (FN) if $H_0$ is not rejected in the test and true positive (TP) if $H_0$ is rejected. 
The true positive rate (TPR) and false positive rate (FPR) used in the computation of the ROC curve are defined as
\begin{align}
    \mathrm{TPR}=\frac{N_\mathrm{TP}}{N_\mathrm{TP}+N_\mathrm{FN}},
    \ \ \ \  \mathrm{FPR}=\frac{N_\mathrm{FP}}{N_\mathrm{FP}+N_\mathrm{TN}},
\end{align}
where $N_\mathrm{A}$, $\mathrm{A} ={\rm TN, FP, FN, TP}$ is 
the number of results classified as $\mathrm{A}$ when the tests are conducted several times. 
FPR is equal to $\alpha$ of the test.

\begin{table}[t]
\centering
\caption{Classification based on the results of the hypothesis test}
\begin{tabular}{llcc}
        &                                   
        & \multicolumn{2}{c}{Null hypothesis} \\
        & \multicolumn{1}{l|}{}             
        & True               
        & False             \\ \cline{2-4} 
\multirow{2}{*}{\rotatebox{90}{Test result}} 
& \multicolumn{1}{l|}{not Significant} 
& 
\begin{tabular}{c}
     True Negative  \\ (TN)
\end{tabular} 
&
\begin{tabular}{c}
     False Negative  \\ (FN)
\end{tabular} 
\\
& \multicolumn{1}{l|}{Significant}       
& 
\begin{tabular}{c}
     False Positive  \\ (FP)
\end{tabular}      &
\begin{tabular}{c}
     True Positive  \\ (TP)
\end{tabular}   
\end{tabular}
\label{table: confusion matrix}
\end{table}

\subsection{Rayleigh statistic}
\label{sec:Rayleigh stat}

The Rayleigh statistic \cite{gwpy, gwpy_doi} is calculated from the time series data as follows. 
First, we set the time resolution, $T$, and divided the data every $T$ seconds. 
The Gaussianity is evaluated with this time resolution.
Let $x(t)$ be one of the divided data, and $x(t)$ is further divided into segments. 
Each segment has a length, $T_\mathrm{FFT}$, and overlaps with the neighboring segment by $T_\mathrm{overlap}$.
$T_\mathrm{overlap}$ was set as $T_\mathrm{FFT}/2$ in this study.
The parameters $T$ and $T_\mathrm{FFT}$ are required the adjustment to fit the non-Gaussianity time scale of interest.
\begin{figure}[t]
    \centering
    \includegraphics[width=140mm]{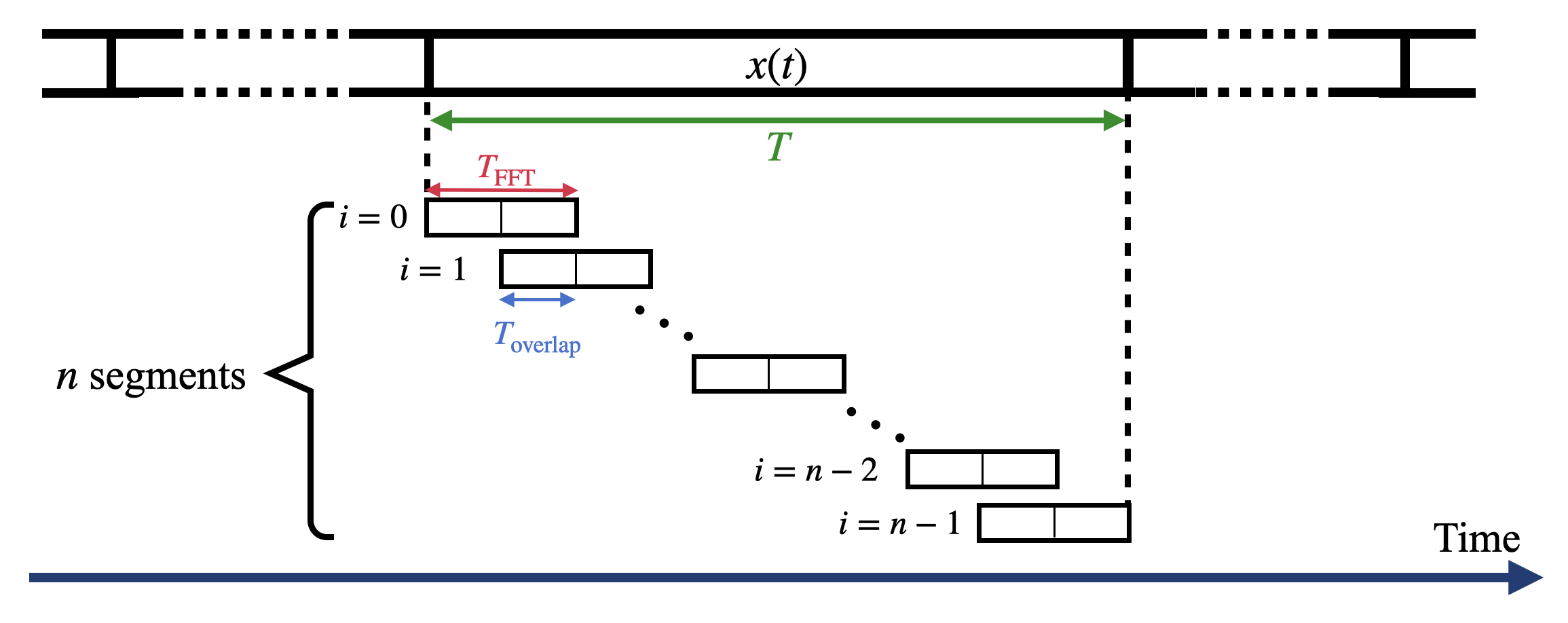}
    \caption{Schematic of time series data segmentation during Rayleigh statistic calculation}
    \label{fig:split timeseries}
\end{figure}
The number of the segments, $n$, is called the sample size and calculated as
$n = 1 + \left\lfloor \frac{T-T_\mathrm{FFT}}{T_\mathrm{FFT}-T_\mathrm{overlap}} \right\rfloor,$
where $\lfloor \cdot \rfloor$ is a floor function.
Figure \ref{fig:split timeseries} shows the schematic view of the above segmentation.
From each segment, a one-sided PSD is defined as
\begin{align}
    P_i(f) = \frac{2 |X_i(f)|^2}{T_\mathrm{FFT}},  \nonumber
\end{align}
where $f$ is the frequency, $i$ is the index of the segment, $i = 0, 1, \cdots n-1$, and
$X_i(f)$ is the frequency spectrum calculated from the $i$-th segment data, $x_i(t)$, defined as
\begin{align}
    X_i(f) = \int_{-\infty}^{\infty} x_i(t)e^{-2\pi ift} dt.
    \label{eq:Fourier}
\end{align}
Then, the Rayleigh statistic, $R(f)$, is defined as
\begin{align}
        R(f)
    = \frac{\sqrt{\displaystyle\frac{1}{n}\sum_{i} (P_i(f)-\overline{P(f)})^2}}{\displaystyle\overline{P(f)} } , 
    \label{eq:RayleighStat}
\end{align}
where 
$\overline{P(f)} = \displaystyle\frac{1}{n}\sum_{i} P_i(f) $.
In other words, $R(f)$ is 
the ratio of the uncorrected sample standard deviation and sample mean of $P_i(f)$.

If $x(t)$ is Gaussian noise, $P_i(f)$ follows an exponential distribution
(for details, see Appendix).
Since the population standard deviation and the population mean of the exponential distribution are equal,
$R(f)$ is expected to be 1.
By contrast,  if data includes non-Gaussian noise,
$P_i(f)$ does not follow an exponential distribution.
Therefore, $R(f)$ does not have to be 1.
For the above reason, the data is likely to be more Gaussian as $R(f)$ is closer to 1 and more non-Gaussian as $R(f)$ is farther from 1.

An expected value of $R(f)$, $E[{R(f)}]$, is biased from 1 depending on $n$, even if $R(f)$ was computed from the true Gaussian noise.
To investigate the degree of the bias, a simple simulation was performed using pseudo-random numbers as the random variables.
We generated $n$ random variables following an exponential distribution and calculated the ratio of the uncorrected sample standard deviation to the sample mean.
Repeating the calculation $10^7$ times, 
we estimated the population mean of $R(f)$ and its error. The error is estimated from the t-distribution with 95\% confidence level.

From this simulation with $n=39$ and 239, we obtained $\overline{R} = 0.964310(88)$ and 0.993826(39), respectively.
We confirmed that even with Gaussian noise, the distribution of the Rayleigh statistic depends on the size of $n$.
As $n$ increases, the deviation from 1 tends to decrease.

\subsection{Extension using hypothesis test}
\label{sec:make test}
In this section, we extend the Rayleigh statistic using the hypothesis test.
First, the background distribution of the Rayleigh statistic will be estimated by the Monte Carlo method.
The background distribution is the distribution of the test statistic when $H_0$ is true and used to calculate the $p$-value.
%

\begin{figure}[t]
    \centering
    \includegraphics[width=100mm]{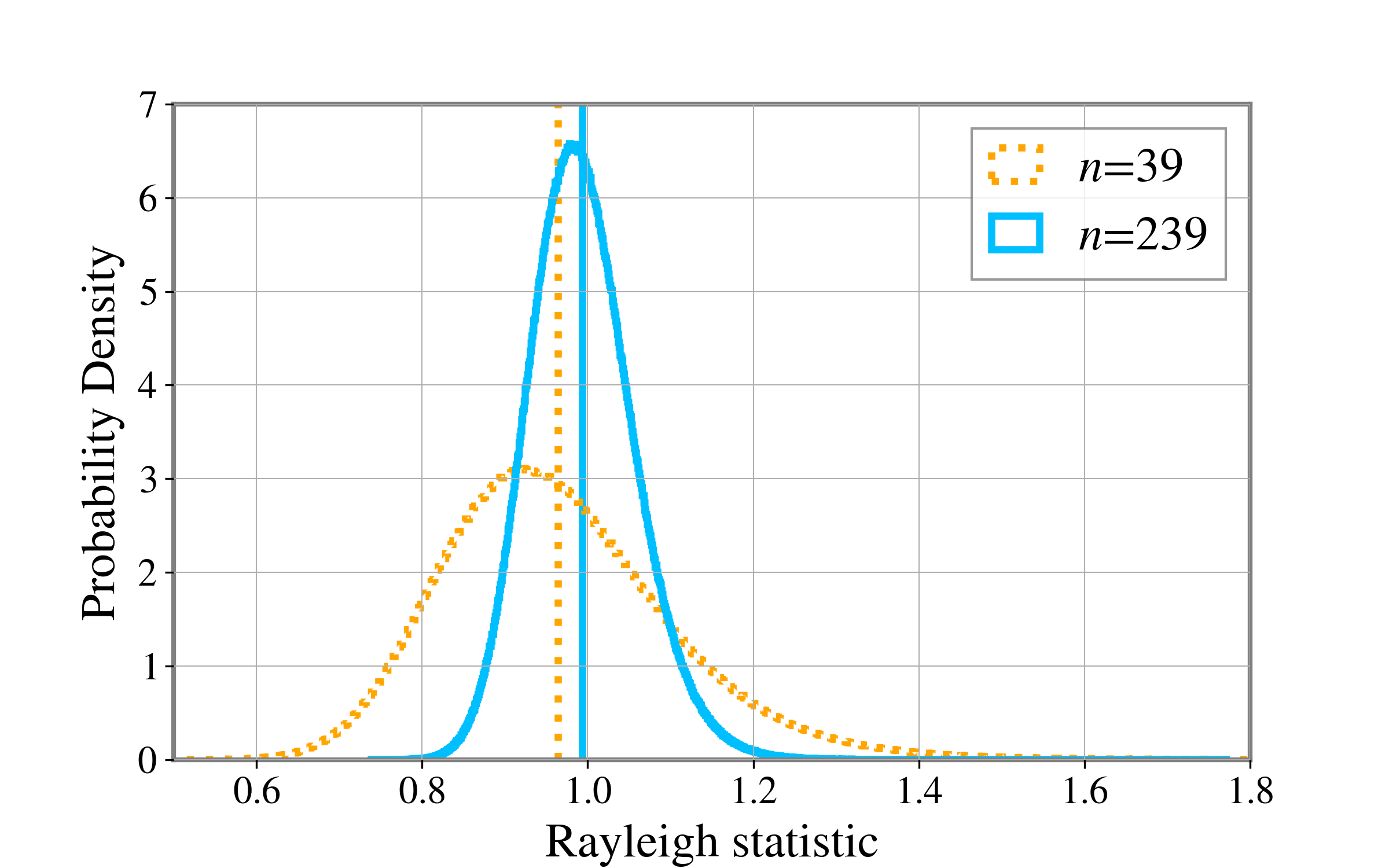}
    \caption{
    Normalized histogram of the background
    distribution including $10^7$ samples for sample sizes $n=39$ and 239.
    Rayleigh statistics were computed from white Gaussian noise.
    The vertical lines
    represent the mean of the distribution.
    For larger $N$,
    the mean value is close to 1.
    These distributions are asymmetric and have a long tail to the right-hand side.
    }
    \label{fig:background_dist}
\end{figure}

Here, we set $H_0$ that the data are Gaussian noise and $H_1$ that the data are non-Gaussian noise. The test statistic, $R$, was set to be the Rayleigh statistic.
To estimate the background distribution by the Monte Carlo method,
we generated white Gaussian noise as the simulated data and computed the Rayleigh statistic for the simulated data with sample size, $n$.
Hereafter, we used 4096 Hz as the sampling rate.
We collected the Rayleigh statistic between 100 and 1900 Hz.
These Rayleigh statistics were treated as independent of time and frequency due to the nature of white Gaussian noise.
Once we simulated and analyzed sufficient data,
we obtained the $10^7$ samples of the Rayleigh statistic following the background distribution.
Figure \ref{fig:background_dist} shows the examples of the background distribution when the sample size is $n=39$ and $239$.
For the following analysis, these background distributions were used.

Considering the data for evaluating the Gaussianity, when $R$ is calculated for the data, the same sample size $n$ as used in the background must be used.
Once the background distribution is estimated,
the $p$-value for $R$ can be computed using Eq. (\ref{eq:p-value}).
If $p < \alpha$, $H_0$ is rejected, and $H_1$ is accepted. In other words, the data is judged as non-Gaussian noise.
We used 0.05 as the significance level, $\alpha$.

\subsection{Validation of the extension}
In this section, we will confirm that the extension performs as expected from the hypothesis test, that is, that the $p$-value follows a uniform distribution.

We validated the results of the extension using the chi-square test.
First, we generated simulated Gaussian noise following KAGRA's design sensitivity using the open-source software, PyCBC \cite{pyCBC, pyCBC_soft}.
From the simulated Gaussian noise, we computed $N_{\rm p} = 10^6$ samples of Rayleigh statistics using Eq. (\ref{eq:RayleighStat}) and $p$-values using Eq. (\ref{eq:p-value}). 

\begin{figure}[t]
    \centering
    \includegraphics[width=100mm]{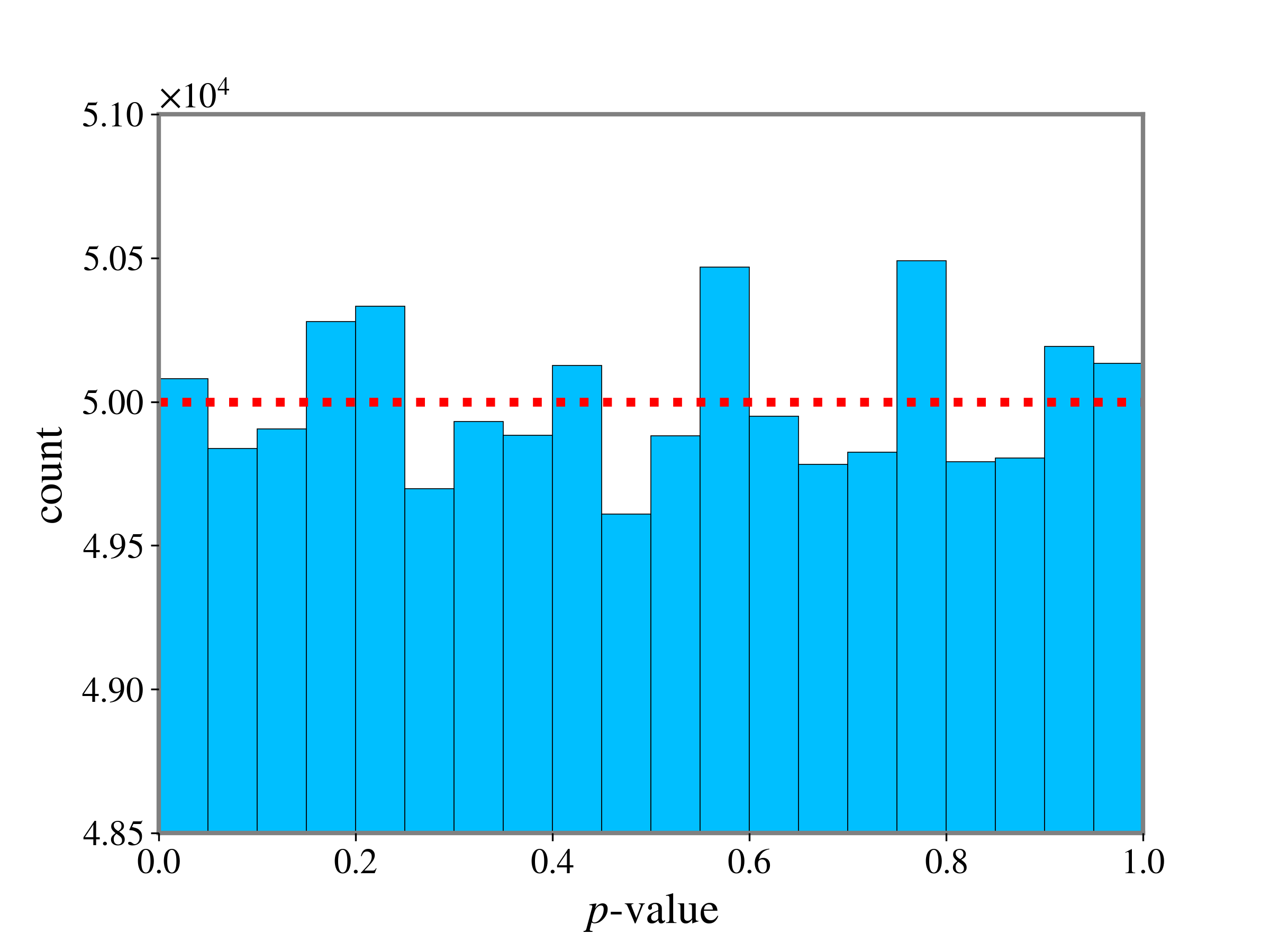}
    \caption{
    Histogram of $p$-values obtained from data following the null hypothesis. Here we set the bin number $N_{\rm bin}=20$ and sample size $n=39$.
    Rayleigh statistics were computed from the simulated Gaussian noise, and $p$-values were obtained from the Rayleigh statistics.
    The red dotted line represents the expected value 
    in the uniform distribution.
    }
    \label{fig:dist. of p-value}
\end{figure}

Figure \ref{fig:dist. of p-value} shows the histogram of the obtained $p$-values in the case of $n=39$.
Since the simulated data follows the null hypothesis,
the $p$-values are expected to follow the uniform distribution.
We tested this using the one-sided chi-square test,
and the test statistic, $\chi^2$, was defined as
\begin{align}
    \chi^2 = \sum_{m=0}^{N_{\rm bin}}{ \frac{( N_m - N_{\rm p}/N_{\rm bin} )^2}{N_{\rm p}/N_{\rm bin}} },\nonumber
\end{align}
where $N_{\rm bin}$ is the number of bins in the histogram, $N_{\rm bin}=20$, and $N_m$ is the number of samples in the $m$-th bin.
We computed $\chi^2$ for the result in Fig. \ref{fig:dist. of p-value}, and obtained $\chi^2 = 24.01$.
The boundary of the rejection region is 30.14 with $\alpha=0.05$, 
and thus, the obtained $\chi^2$ is outside the rejection region.
We concluded that the distribution of $p$-values
is not significantly different from the uniform distribution and confirmed that the extension satisfies the hypothesis test.

\section{Evaluation of detection efficiency}
\label{sec:evaluation}
The hypothesis test for the Rayleigh statistic makes it possible to infer that the data is non-Gaussian noise with given $\alpha$.
By assuming a non-Gaussian noise model, the detection efficiency for that model can be evaluated.
In this section, we discuss the detection efficiency of the test against two non-Gaussian models and the effects of $n$ and $T_\mathrm{FFT}$ on the detection efficiency.

\subsection{Models of non-Gaussian noise}
\label{sec: modeling}
Non-Gaussian noise can be divided into non-stationary noise, which appears only for short periods, and stationary noise, which exists regardless of time.
Here, we considered two well-modeled stationary and non-stationary non-Gaussian noise profiles.

\subsubsection{model I : Gaussian noise superposition}
Model I reproduces non-stationary noise with transient Gaussian noise within stationary Gaussian noise.
The time series data, $x(t)$, following the noise model I can be expressed as
\begin{align}
    x(t) = n_0(t) + A_1 \cdot B(t) n_1(t),\nonumber
\end{align}
where 
$n_0$ and $n_1$ are independent Gaussian noise with lengths $T$ generated by the method described in Section \ref{sec:make test},
$A_1$ is a factor
to control the degree of non-Gaussianity, 
and $B(t)$ is a window function for the noise component to be smoothly connected.
We used the Tukey window for $B(t)$ as
\begin{align}
    B(t) = 
    \begin{cases}
    \mathrm{Tukey}(\beta = 0.5) &  (t_0 < t < t_0 + t') \\ \nonumber
    0 &  (otherwise.),
    \end{cases}
\end{align}
where $t_0$ and $t'$ are the onset time and duration of the transient noise, respectively; and $\beta$ is the shape parameter of the Tukey window, which matches the Hann window when $\beta=1$ and matches the Rectangular window when $\beta=0$.
In this simulation, the duration of the transient noise was fixed as $t'=T/6$ sec, and $t_0$ was a random integer satisfying $0 \le t_0 \le T-t'$.

\begin{figure}[t]
        \begin{minipage}[b]{0.5\linewidth}
        \centering
        \includegraphics[width=0.95\linewidth]{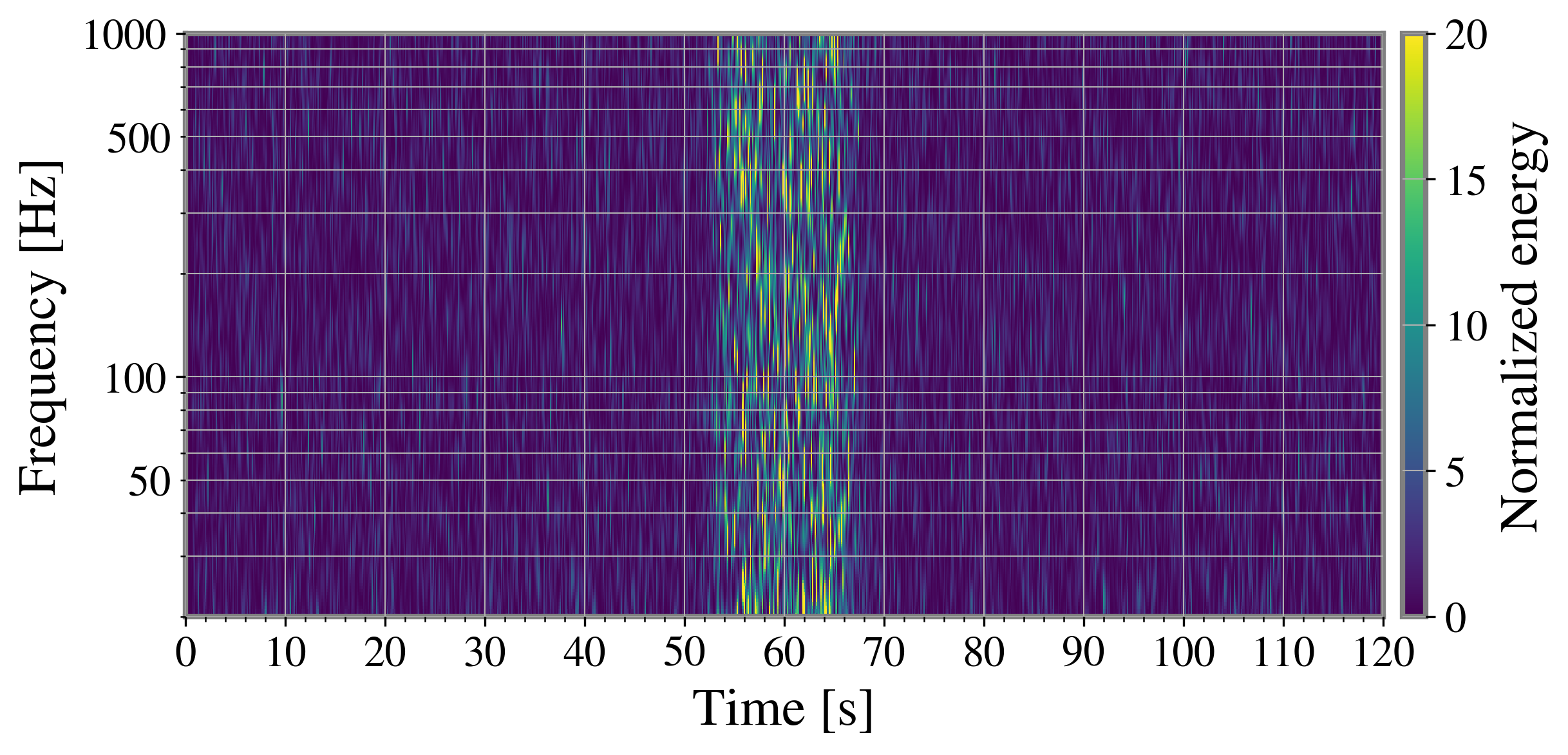}
        \label{fig:model1_qtrans}
        \end{minipage} 
        \begin{minipage}[b]{0.5\linewidth}
        \centering
        \includegraphics[width=0.95\linewidth]{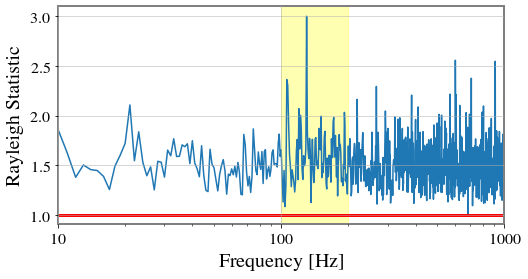}
        \label{fig:model1_rayleighspectrum}
        \end{minipage} 
        \caption{(Left) Q transform of data generated according to the model I ($A_1=2$), the color bar represents normalized energy.
        (Right) Rayleigh statistic $R(f)$ estimated from the generated data, where $T_{\rm FFT}=1$ s. The red line is the expected value of the Rayleigh statistic when the sample size is sufficiently large. The yellow band is the frequency band selected when drawing the ROC curve (100 $\sim$ 200 Hz).}
        \label{fig:model1_plots}
\end{figure}

Figure \ref{fig:model1_plots} shows the Q transform \cite{gwpy, gwpy_doi,qtransform} and its Rayleigh statistic at each frequency for the simulated noise. 
Q transform is a signal processing method that transforms time series data into the frequency domain and visualizes the data.
Figure \ref{fig:model1_plots} shows that the normalized energy of the Q transform increases in all frequency bands in the presence of transient noise.
The Rayleigh statistic also increases in all frequency bands compared to the expected value, as indicated by the red line.

\subsubsection{model \two : scattered light noise}
Model \two\ reproduces the scattered light noise that have limited the sensitivity in LIGO and Virgo \cite{scatter_virgo,scatter_LIGO}. 
This noise appears as a change in the phase of the propagating light as the distance between the scatterer and mirror fluctuates.
The scattered light noise can be considered stationary noise if the duration of the noise is longer than the one we focus on.
The amplitude of the scattered light noise in strain data, $h_\mathrm{sc}$, is modeled in \cite{scatter_virgo} as
\begin{align}
    h_\mathrm{sc}(t) = G \cdot \sin \left( \frac{4\pi}{\lambda} (x_0+\delta x_{sc}(t))\right),\nonumber
\end{align}
where $G$ is a coupling factor, $\lambda$ is the optical wavelength of the laser, $x_0$ is the static distance between the scatterer and mirror, and $\delta x_\mathrm{sc}$ is the fluctuation along the direction of beam.

Here, we considered the case of a single bounce between the scatterer and mirror.
Specifically, the fluctuation $\delta x_\mathrm{sc}$ is written as
\begin{align}
    \delta x_{\rm sc} = a_2 \cos (2\pi f_{\rm sc} t). \nonumber
\end{align}
where $a_2$ and $f_{\rm sc}$ are respectively the amplitude and frequency of the single bounce.
We fixed $\lambda=1064$ nm, $x_0 = 0$ m, $G = 3\times 10^{-22}$, and $f_{\rm sc}=0.2 \ \mathrm{Hz}$.
This $f_{\rm sc}$ is the typical frequency of ground vibrations such as that due to microseism.
To mimic the scattered light noise in \cite{scatter_LIGO}, we modulated $a_2$  to have a 25\% fluctuation centered on the constant factor, $A_2$,
which controls the degree of non-Gaussianity.

The time series data, $x(t)$, following the noise model \two \  can be expressed as
\begin{align}
    x(t) = n_0(t) + h_{\rm sc}(t).
    \label{eq:scatter=gauss+hsc}
\end{align}

\begin{figure}[t]
        \begin{minipage}[b]{0.5\linewidth}
        \centering
        \includegraphics[width=0.95\linewidth]{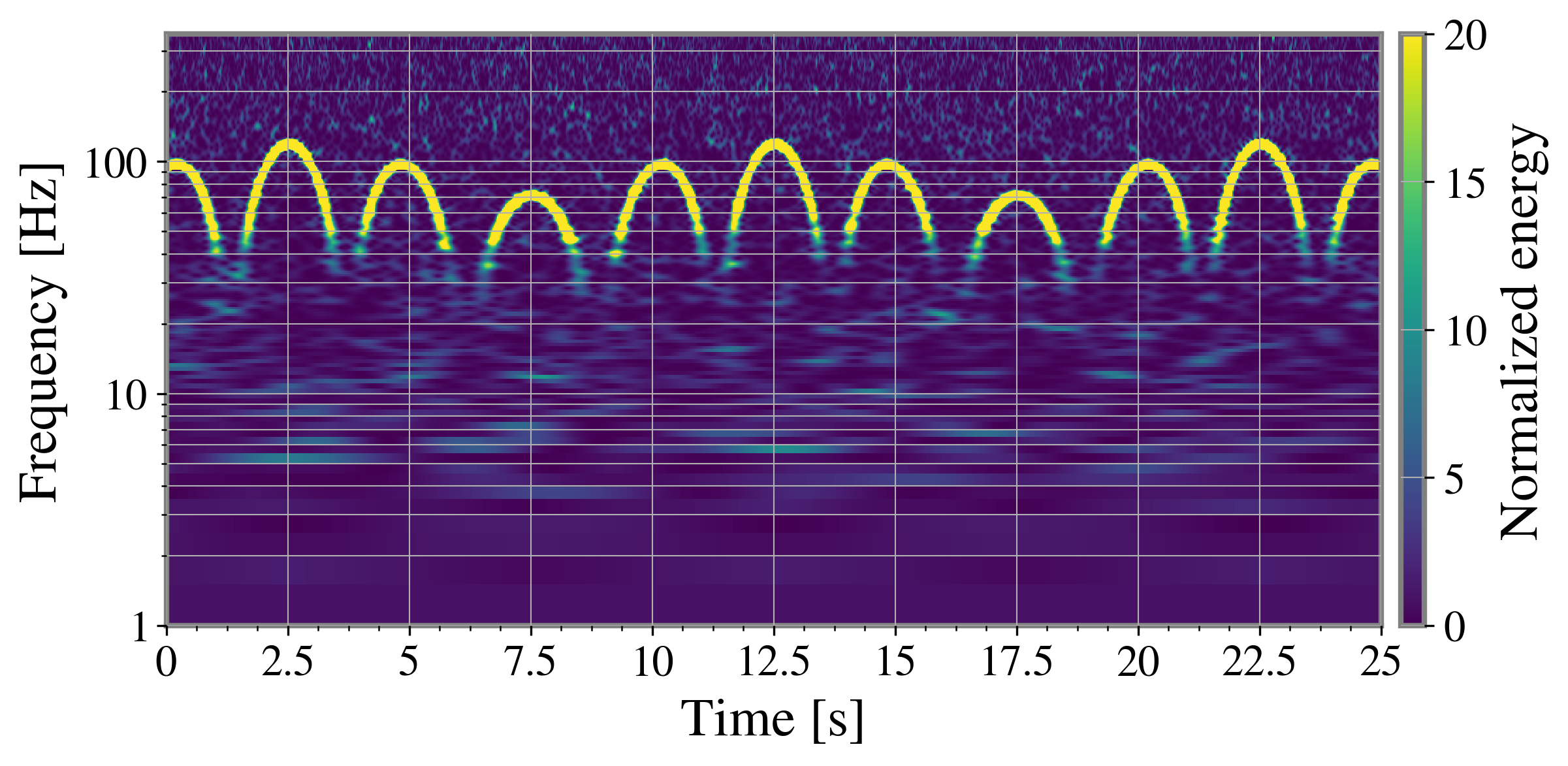}
        \end{minipage} 
        \begin{minipage}[b]{0.5\linewidth}
        \centering
        \includegraphics[width=0.95\linewidth]{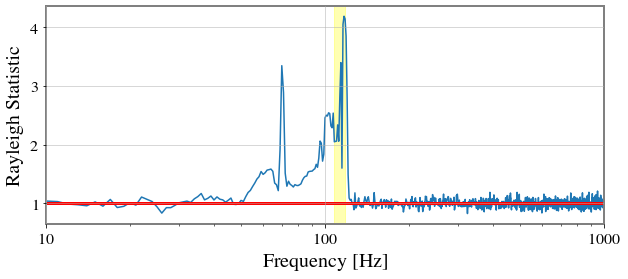}
        \end{minipage} 
        \caption{(Left) Q transform of data generated according to the model \two \ ($A_2=4\times 10^{-5}$). The color bar represents the normalized energy.
        (Right) Rayleigh statistic $R(f)$ estimated from the generated data, where $T_{\rm FFT}=1$ s. The red line is the expected value of the Rayleigh statistic when the sample size is sufficiently large. The yellow band is the frequency band selected when drawing the ROC curve (108 $\sim$ 118 Hz) }
        \label{fig:model2_plots}
\end{figure}

Figure \ref{fig:model2_plots} shows the Q transform of $x(t)$.
The duration lasting two arches corresponds to a period of  oscillation of the scatterer, that is, $1/f_\mathrm{sc}$.
The frequency of the oscillation $f_\mathrm{sc}$ is 0.2 Hz, but in the Q transform, the noise appears at 50$\sim$100 Hz.
It is known that the frequency, $f_{\rm fringe}$, at which the scattered light noise appears in the time-frequency map can be written as \cite{scatter_virgo}
\begin{align}
    f_{\rm fringe}(t) =\left| \frac{2 v_{\rm sc}(t)}{\lambda} \right|,
    \label{eq:f_fringe}
\end{align}
where $v_{\rm sc}(t)$ is the velocity of the scatterer.
In our model, the maximum frequency, $f_{\rm max}$, of $f_{\rm fringe}$ is
\begin{align}
    f_{\rm max} \simeq 1.25A_2 \times \frac{4\pi f_{\rm sc}}{\lambda}.
    \label{eq:freq_max}
\end{align}

Figure \ref{fig:model2_plots} shows an example of the Rayleigh statistics at each frequency.
In the frequency band affected by scattered light noise, the Rayleigh statistics show large excesses.

\subsection{Results}
Using the noise described in the previous section, we evaluated the detection efficiency of the hypothesis-tested Rayleigh statistic.
We computed the ROC curve and its area under the curve (AUC).
The ROC curve is a plot of FPR versus TPR when the threshold for rejection is changed by sweeping the FPR between 0 and 1.
AUC is the area enclosed by TPR$=0$, FPR$=1$, and the ROC curve.
When the AUC is large, the detection capability is high.
The AUC is distributed between 0 and 1 by definition.
In the case of random detection, the ROC curve is a straight line connecting (0,0) and (1,1), with AUC=0.5.

We generated noise data
following the models I and \two.
We used three parameter sets to mimic the different degrees of non-Gaussianity,
and referred to them as weak, mild, and strong.
Table \ref{table: parameter} shows these parameter sets.
To investigate the effect of the sample size $n$ and $T_\mathrm{FFT}$ on the detection efficiency,
we performed the simulations with changing $T$ and $T_\mathrm{FFT}$.

In drawing the ROC curve,
the Rayleigh statistic which is affected by non-Gaussian noise, should be used.
Figures 4 and 5 show that non-Gaussian noise spreads in the frequency region.
Therefore, the detection efficiency is estimated with an appropriate frequency band that includes non-Gaussian noise rather than only a single frequency.
If the selected frequency band is $[f_1, f_2]$, the corresponding frequency range is $\Delta f = f_2 - f_1$.
The number of Rayleigh statistics obtained in this frequency band in one simulation is $\Delta f \times T_\mathrm{FFT}$.
By $N$ iterations, $N \times \Delta f \times T_\mathrm{FFT}$ samples of Rayleigh statistics are obtained.
Here, we used $N=1000$.
With each Rayleigh statistic, the $p$-value and TPR were calculated.
From the Rayleigh statistics in Fig. \ref{fig:model1_plots}, we selected the frequency band $[f_1, f_2]=[100,200]$ for model I.
From the Rayleigh statistics in Fig. \ref{fig:model2_plots}, we selected the frequency band $[f_1, f_2]=[f_{\rm max}-10, f_{\rm max}]$ for model \two.

The frequency band set here can be selected as long as it contains noise.
The TPR of the calculated ROC curve represents the proportion of rejections among the entire Rayleigh statistic within that frequency band.
When the Rayleigh statistic is independent of frequency, as in model I, the TPR does not change regardless of the frequency band selected.
However, when the Rayleigh statistic has a frequency dependence as in model \two, the TPR depends on the selected frequency band.
This is because the distribution of the Rayleigh statistic is different for each frequency, and the detection efficiency also changes,
for example, when there is a frequency with a particularly high detection efficiency. 
In this case, the TPR will be higher if the frequency is included in the selected frequency band and lower if not.
Here, 10 Hz was selected as the frequency range with high power in the Q transform, as shown in Fig. 5.
By contrast, since FPR is a quantity determined from the background distribution, it does not depend on the noise model but only on $\alpha$.

\begin{figure}[t]
    \centering
    \includegraphics[width=140mm]{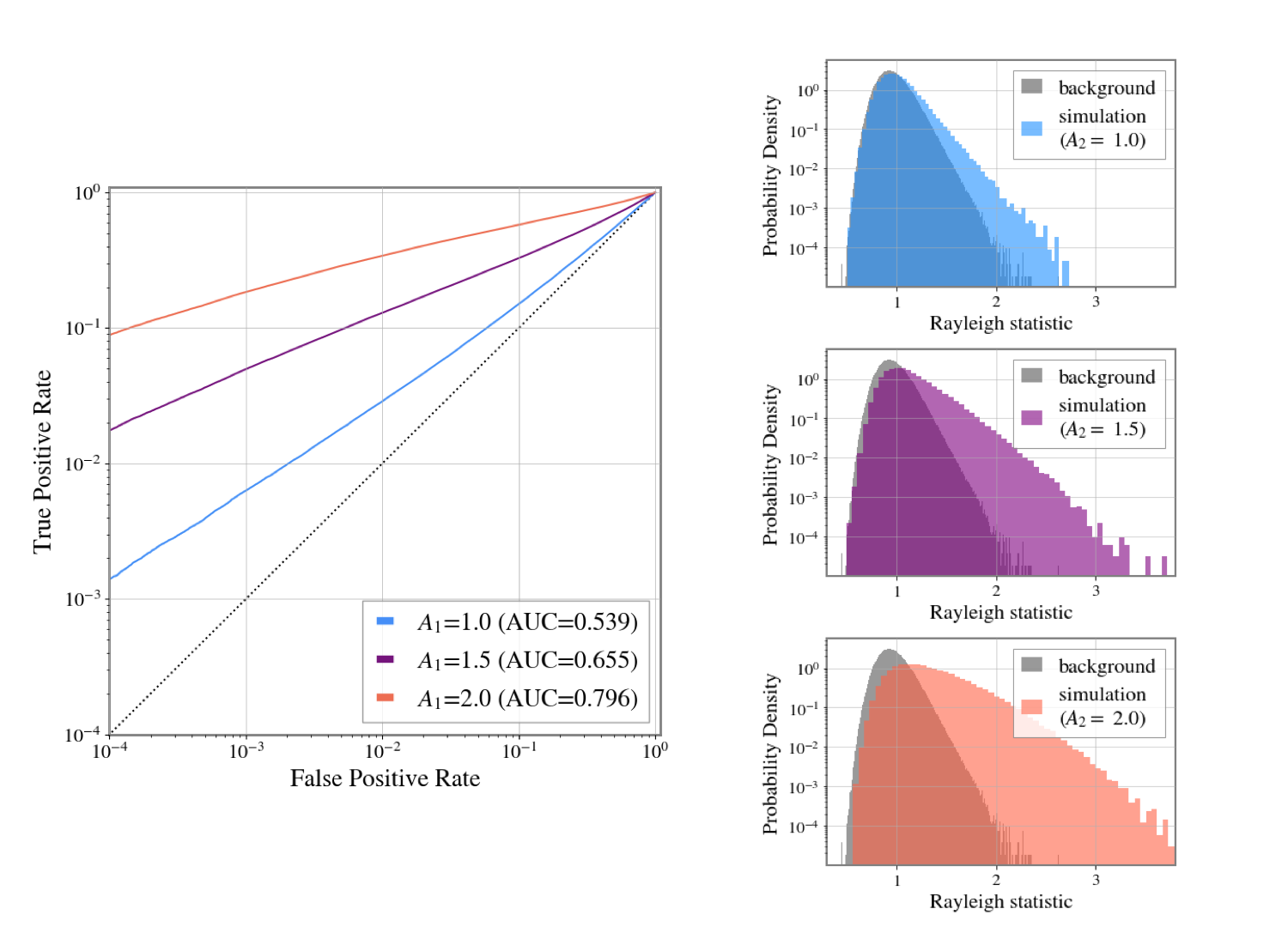}
    \caption{
    Simulation with $T=120 \mathrm{\ s},\  T_{\rm FFT}=6 \mathrm{\ s}$, and $n=39$ in model I. (Left) ROC curves.
    Blue, purple, orange curves represent the ROC curve with parameters of the weak, mild, strong cases, respectively.
    (Right) Rayleigh statistic distribution in the simulation and the background distribution.
    }
    \label{fig:model1_120_6}
\end{figure}

\begin{figure}[t]
    \centering
    \includegraphics[width=140mm]{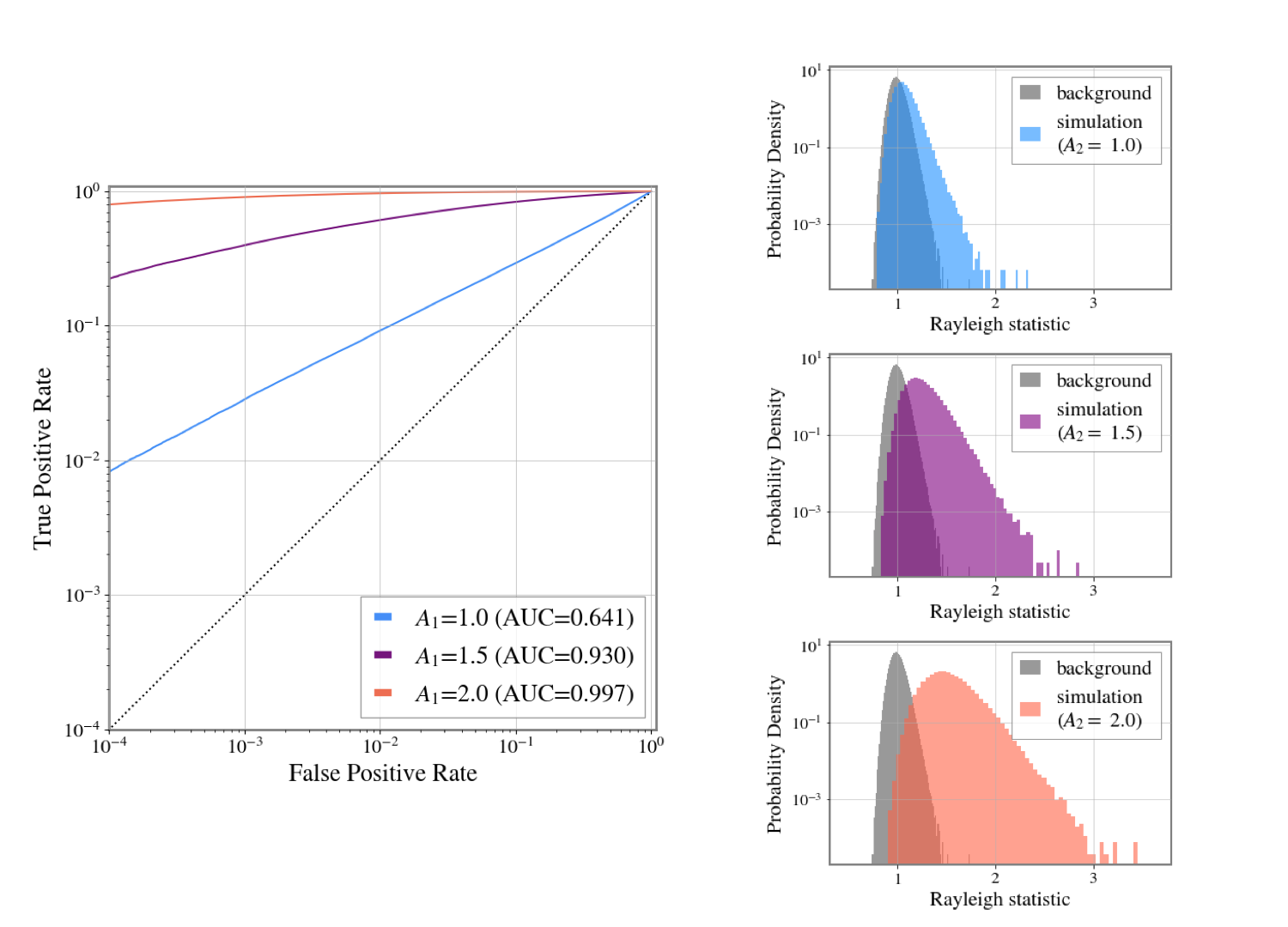}
    \caption{
    Simulation with $T=720 \mathrm{\ s},\  T_{\rm FFT}=6 \mathrm{\ s}$, and $n=239$ in model I. (Left) ROC curves. 
    Blue, purple, orange curves represent the ROC curve with parameters of the weak, mild, strong cases, respectively.
    (Right) Rayleigh statistic distribution in the simulation and the background distribution.
    }
    \label{fig:model1_720_6}
\end{figure}

\begin{figure}[t]
    \centering
    \includegraphics[width=140mm]{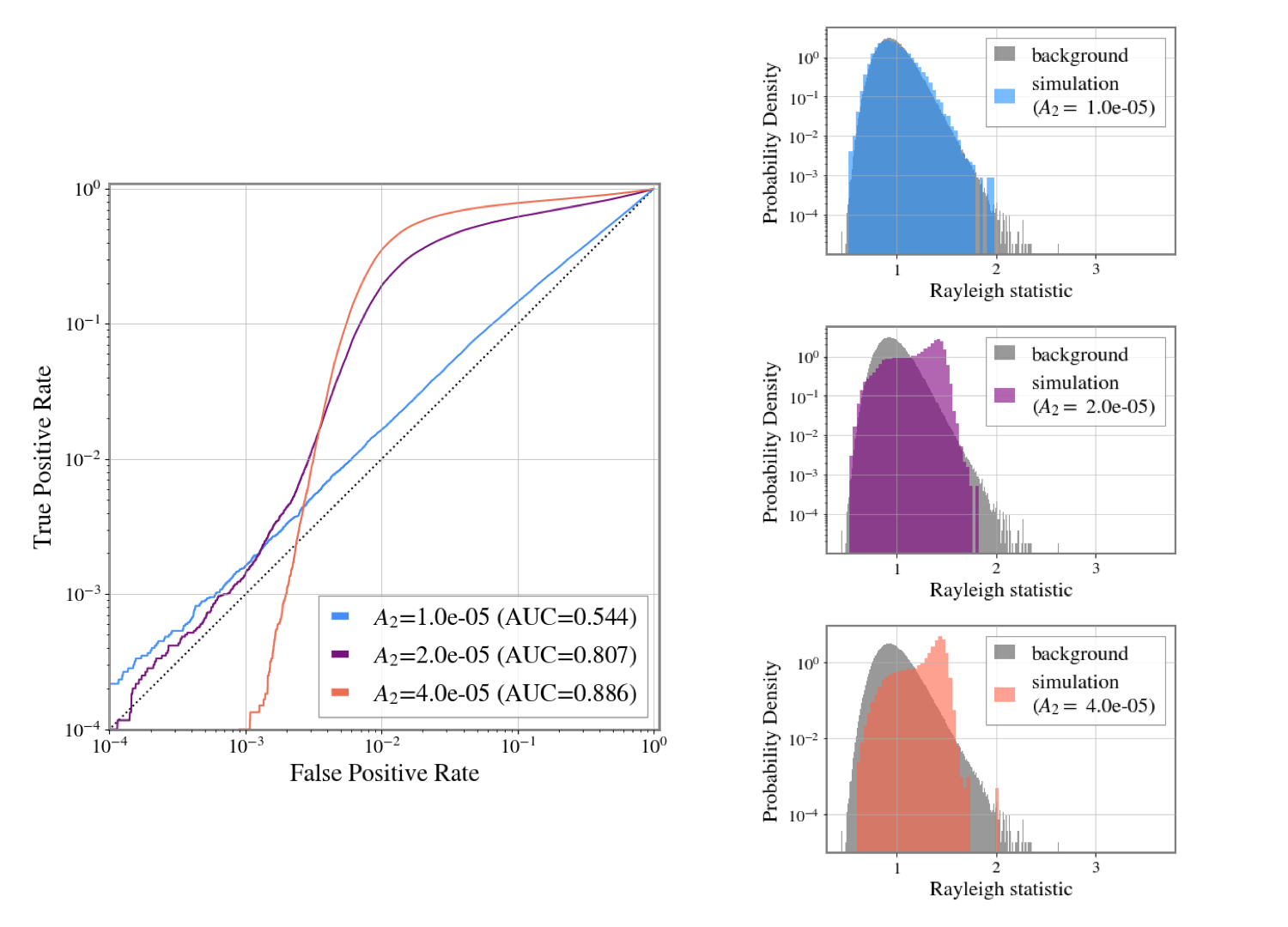}
    \caption{
    Simulation with $T=120 \mathrm{\ s},\  T_{\rm FFT}=6 \mathrm{\ s}$, and $n=39$ in model \two . (Left) ROC curves. 
    Blue, purple, orange curves represent the ROC curve with parameters of the weak, mild, strong cases, respectively.
    (Right) Rayleigh statistic distribution in the simulation and the background distribution.
    }
    \label{fig:model2_120_6}
\end{figure}

\begin{figure}[t]
    \centering
    \includegraphics[width=140mm]{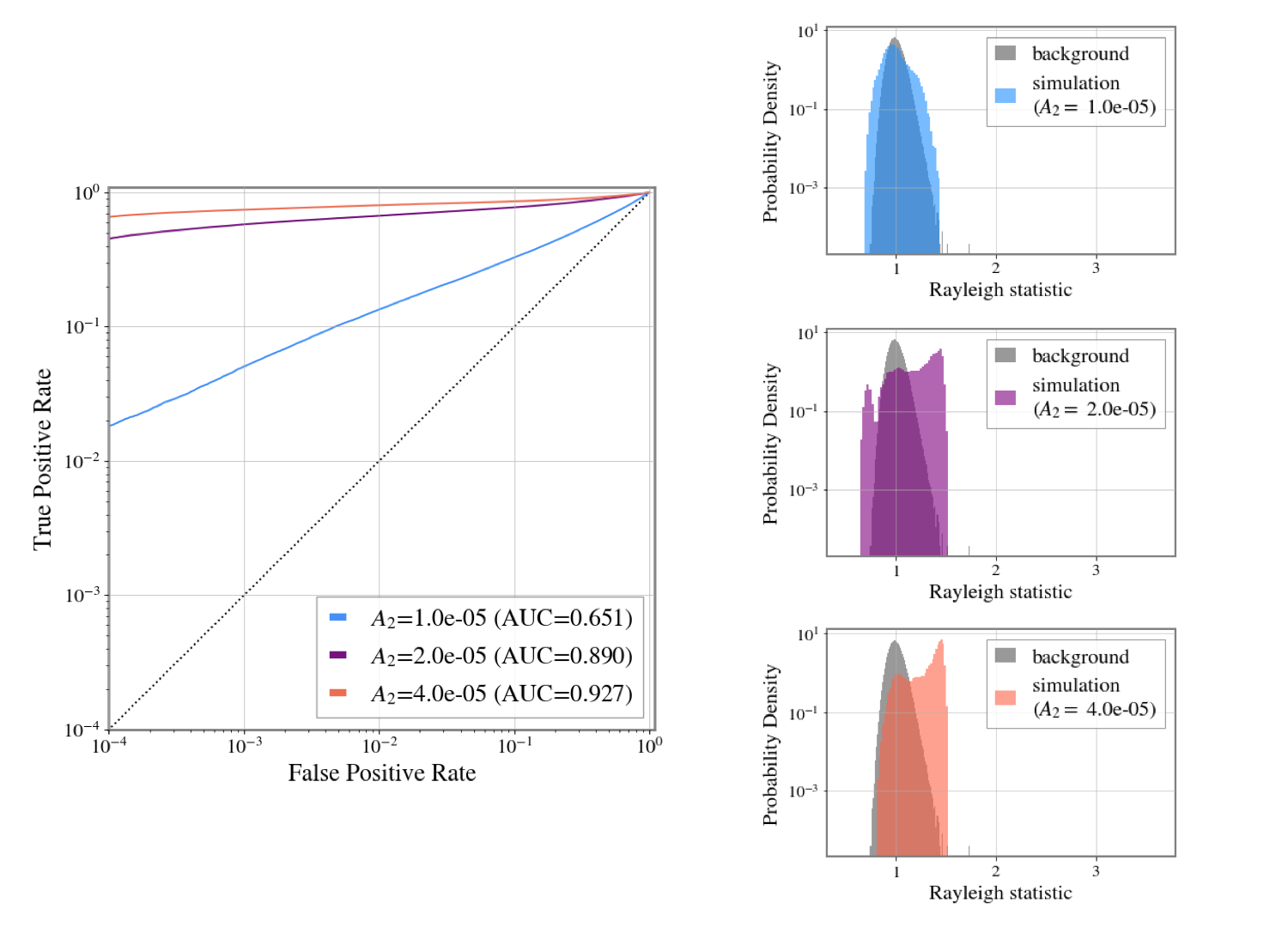}
    \caption{
    Simulation with $T=720 \mathrm{\ s},\  T_{\rm FFT}=6 \mathrm{\ s}$, and $n=239$ in model \two . (Left) ROC curves. 
    Blue, purple, orange curves represent the ROC curve with parameters of the weak, mild, strong cases, respectively.
    (Right) Rayleigh statistic distribution in the simulation and the background distribution.
    }
    \label{fig:model2_720_6}
\end{figure}

\begin{figure}[t]
    \centering
    \includegraphics[width=140mm]{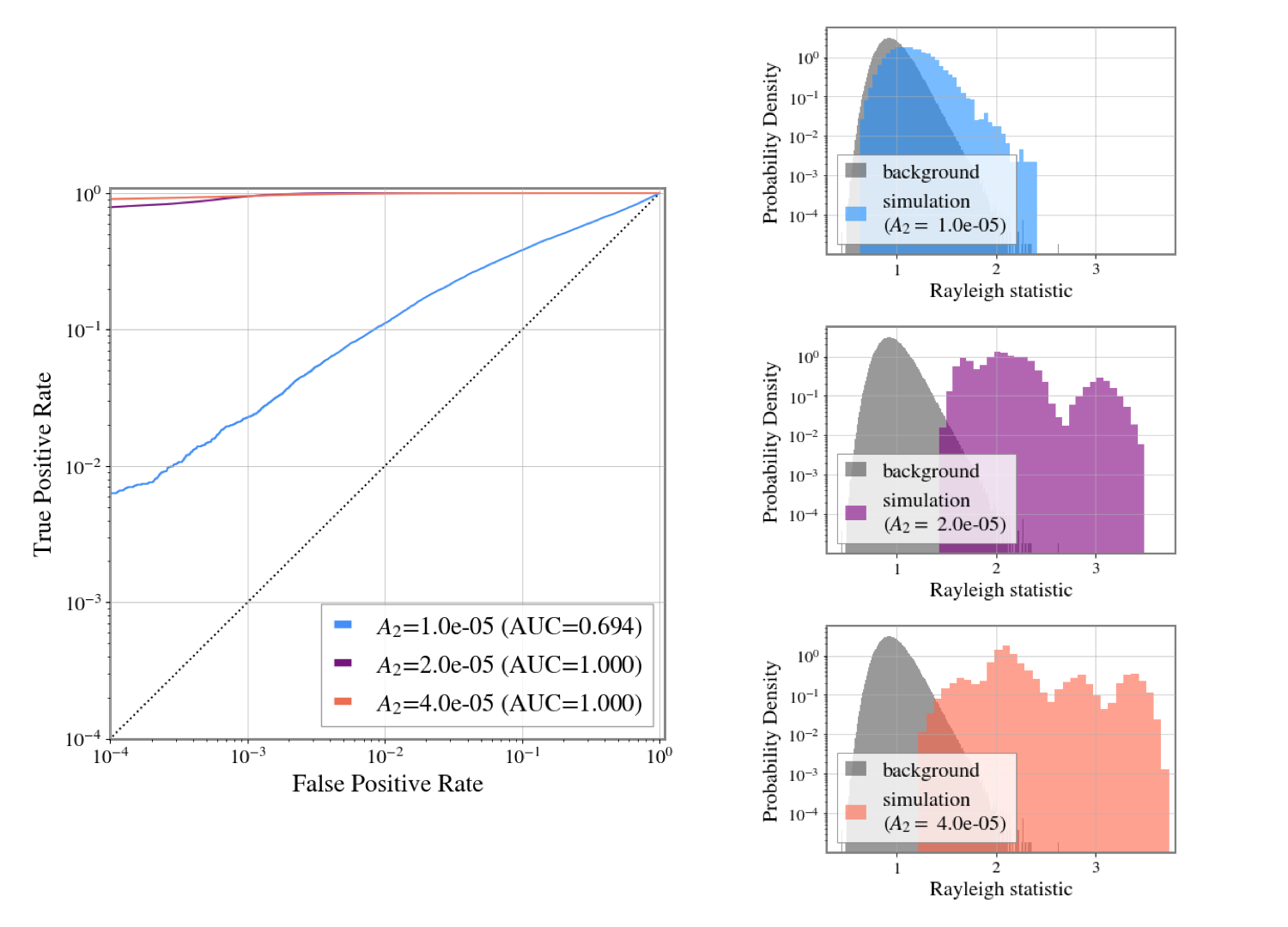}
    \caption{
    Simulation with $T=20 \mathrm{\ s},\  T_{\rm FFT}=1 \mathrm{\ s}$, and $n=39$ in model \two . (Left) ROC curves. 
    Blue, purple, orange curves represent the ROC curve with parameters of the weak, mild, strong cases, respectively.
    (Right) Rayleigh statistic distribution in the simulation and the background distribution.
    }
    \label{fig:model2_20_1}
\end{figure}

Figures \ref{fig:model1_120_6} and \ref{fig:model1_720_6} show the ROC curves and distributions of the Rayleigh statistic for model I with $n=39$ and $n=239$, and $T_{\rm FFT}=6 \mathrm{\ s}$. 
Figures \ref{fig:model2_120_6}, \ref{fig:model2_720_6}, and \ref{fig:model2_20_1} show the ROC curves and distributions of the Rayleigh statistic for model \two with $n=39$ and $T_{\rm FFT}=6 \mathrm{\ s}$, $n=239$ and $T_{\rm FFT}=6 \mathrm{\ s}$, and $n=39$ and $T_{\rm FFT}=1 \mathrm{\ s}$, respectively. 

\begin{table}[t]
\centering
\caption{Parameters used for making ROC curves}

\begin{tabular}{c|c|ccc}
model  & parameter & weak  & mild  & strong \\ \hline
model I & $A_1$ & $1$ & $1.5$ & $2$\\
model \two & $A_2$ & $1\times 10^{-5}$   & $2\times 10^{-5}$   & $4\times 10^{-5}$

\end{tabular}
\label{table: parameter}
\end{table}


\subsection{Discussion}
In this section, we discuss the ROC curve from simulated non-Gaussian noise.

In both of Figs. \ref{fig:model1_120_6} and \ref{fig:model1_720_6}, 
detection efficiency increased as the degree of non-Gaussianity increased from weak to strong.
The distributions of the Rayleigh statistic from simulations show that a higher Rayleigh statistic was obtained as the degree of the non-Gaussianity increased from weak to strong.
In Fig. \ref{fig:model1_120_6} left, the AUC increased from weak to strong, with values of 0.539, 0.655, and 0.796.
Comparing Fig. \ref{fig:model1_120_6} and Fig. \ref{fig:model1_720_6},
the sample size was increased to $n=239$.
Here, the AUCs from weak to strong were 0.641, 0.930, and 0.997. 
These AUCs are larger than the those of Fig. \ref{fig:model1_120_6}.
The larger sample size reduced the variance of the Rayleigh statistic distribution in both the background and simulation, which increased the detection efficiency.

In the right part of Figs. \ref{fig:model2_120_6}, \ref{fig:model2_720_6}, and \ref{fig:model2_20_1}, 
the characteristic shape with multiple peaks in the distribution of the Rayleigh statistic can be observed, which is caused by the frequency dependence of the Rayleigh statistic.
Figure \ref{fig:model2_plots} shows the frequency dependence of the Rayleigh statistic. 
It can be seen that a finer frequency dependence exists even within the frequency band of 10 Hz, which was selected when drawing the ROC curve.
Therefore, the distribution of the Rayleigh statistic was different for each frequency within 10 Hz, and their superposition resulted in a structure with multiple peaks.

In Fig. \ref{fig:model2_120_6},
as for the mild and strong cases, the TPR is above 0.1 for FPR$>10^{-2}$.
TPR decreases sharply for FPR$<10^{-2}$, which indicates worse performance than that with random detection.

Comparing Figs. \ref{fig:model2_720_6} and \ref{fig:model2_120_6}, the sample size $n$ in Fig. \ref{fig:model2_720_6} was increased to $n=239$ with the same $T_\mathrm{FFT}$.
Here, all ROC curves were above the random detection line.
As the sample size increased, the detection efficiency increased.

Comparing Fig. \ref{fig:model2_20_1} and Fig. \ref{fig:model2_120_6}, the FFT length $T_\mathrm{FFT}$ of Fig. \ref{fig:model2_20_1} was reduced to $T_\mathrm{FFT} = 1$ with the same $n$.
The detection efficiency improved for all 
the three degrees of non-Gaussianity.
Particularly, for the mild and strong cases, AUC=1.000, which is approximately 100\% of the detection efficiency.
In this simulation, as in the other cases, $f_\mathrm{sc}=0.2 \ \mathrm{Hz}$ was used, so the period of the arch structure of the scattered light noise is 2.5 sec, as seen in the left panel of Fig. \ref{fig:model2_plots}.
If $T_\mathrm{FFT}$ is lower than this period, the distribution of the Rayleigh statistic from the simulation is likely to deviate from that of the background.
As a result, a higher detection efficiency is obtained.

In the case of non-stationary noise, such as in model I, the Rayleigh statistic tends to be large compared to that for the background distribution.
This is because a part of PSDs affected by non-Gaussian noise have larger values than others, resulting in a large standard deviation of the PSD distribution.
By contrast, in the case of stationary noise, such as in model \two, the noise is likely to dominate the data at a specific frequency. Therefore, the Rayleigh statistic depends on the noise model and parameters used in the analysis.

\begin{figure}[t]
        \begin{minipage}[b]{0.5\linewidth}
        \centering
        \includegraphics[width=0.95\linewidth]{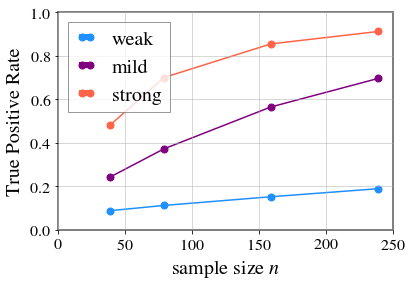}
        \end{minipage} 
        \begin{minipage}[b]{0.5\linewidth}
        \centering
        \includegraphics[width=0.95\linewidth]{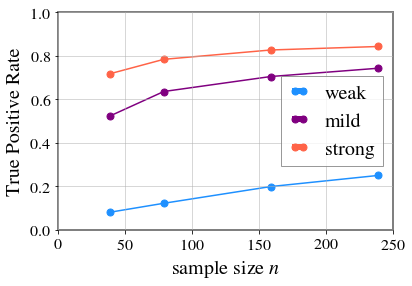}
        \end{minipage} 
        \caption{
        (Left)The relationship between sample size and TPR at the significance level $\alpha=0.05$ for data following model I, where $T_{\mathrm{FFT}}=6$ s. The sample sizes are n=39, 79, 159, 239. 
        (Right) The relationship between the sample size and true positive rate at the significance level $\alpha=0.05$ for data following model \two, where $T_{\mathrm{FFT}}=6$ s. The sample sizes are n=39, 79, 159, 239. 
        }
        \label{fig:power}
\end{figure}

As shown in the comparison between Figs. \ref{fig:model1_120_6} and \ref{fig:model1_720_6}, and between Figs. \ref{fig:model2_120_6} and \ref{fig:model2_720_6}, 
it is important to discuss the relationship between the sample size and detection efficiency.
We examined the detection efficiency for each model and degree of non-Gaussian noise when the sample size was varied as $n=39,79,159,239$ with a fixed significance level $\alpha=0.05$.
We fixed $T_\mathrm{FFT}=6\ \mathrm{s}$.
To change $n$, we changed the data length, $T$.
Figure \ref{fig:power} shows the relationship between $n$ and the TPR in the two models.
For both models, the efficiency increased as the sample size increased.
In the case of model I, for the mild and strong cases, the efficiency was relatively high when the sample size increased.
For the strong noise model, the TPR was from 0.482 to 0.912.
In model \two, even for the mild case, the TPR was over 0.5 for $n=39$, confirming that more than half of the signal can be detected for these models and parameters.
For the strong noise model, the TPR ranged from 0.718 to 0.843.

The sample size or $T_\mathrm{FFT}$ can be adjusted to the non-Gaussian noise of interest or required detection efficiency.

\section{Conclusions}
Noise in data from GW detector is classified into two categories, Gaussian noise and non-Gaussian noise.
Here, we focused on non-Gaussian noise, which affects GW analysis because many analysis methods are optimized for Gaussian noise.
There have been attempts to characterize non-Gaussian noise. 
We used the Rayleigh statistic, which can quantitatively evaluate the Gaussianity.
However, it was difficult to set a threshold that determines if the data is non-Gaussian noise.
We proposed an extension of the Rayleigh statistic by introducing a hypothesis test.
The null hypothesis that the data was Gaussian noise was set.
The $p$-value was calculated by comparing the background distribution with the Rayleigh statistic.
The background distribution was estimated by the Monte Carlo method.
When the $p$-value was smaller than the significance level, $\alpha$, we judged the data is non-Gaussian noise.
When the null hypothesis was true, the extension was validated by confirming that the distribution of the $p$-value was not significantly different from the uniform distribution.

The extension allows us to evaluate the detection efficiency against a specific noise model with given $\alpha$.
In this study, we considered two models: stationary and non-stationary noise.
The simulated noise following each model were generated,
and the ROC curve was estimated.
Model I reproduces non-stationary
noise with transient Gaussian noise within stationary Gaussian noise,
and model \two \ reproduces the scattered light noise.

For model I, the simulated noise was stably determined to be non-Gaussian noise.
This is because the value of the Rayleigh statistic increased over the entire frequency.
The detection efficiency increased as the sample size increased.
For model \two,
the Rayleigh statistics depended on the frequency, resulting in the characteristic shapes with multiple peaks in the distributions. 
We found that the ROC curves differed significantly when the sample size and $T_\mathrm{FFT}$ were changed.
We also examined the relationship between the sample size and detection efficiency when the significance level was 0.05.
In both models, the detection efficiency increased with increasing sample size.
Focusing on the strong scattered light noise model, the TPR was always above 0.7.

These simulation confirmed that detection is possible for both models, although the Rayleigh statistic does not assume a noise model in the calculations.
Additionally, the extension requires low computational time once the background distribution is estimated, because the judgement can be made by comparing the Rayleigh statistic with the threshold.
Therefore, the extension can serve as a new indicator for initial detection of non-Gaussian noise.
It is also expected to be applied to Virgo's rayleighSpectro and other analysis methods.


\section*{Acknowledgment}

S.Y. and H.Y. thank Shuhei Mano for fruitful discussions.
This work was carried out by the joint research program of the Institute for Cosmic Ray Research (ICRR), The University of Tokyo.
This work was supported by JSPS KAKENHI Grant Number JP23K13120.

Various parts of the analysis used GWpy \cite{gwpy,gwpy_doi} and PyCBC \cite{pyCBC,pyCBC_soft}.


%

\vspace{0.2cm}
\noindent









\section*{References}
\let\doi\relax
\bibliographystyle{ptephy}
\bibliography{reference}

\begin{thebibliography}{10}

\bibitem{PhysRevLett.116.061102}
B.~P. Abbott et~al., Phys. Rev. Lett., {\bf 116}, 061102 (Feb 2016).

\bibitem{Abbott_2016}
B.~P. Abbott et~al., The Astrophysical Journal Letters, {\bf 818}(2), L22 (feb 2016).

\bibitem{aLIGO}
Gregory~M Harry et~al., Classical and Quantum Gravity, {\bf 27}(8), 084006 (apr 2010).

\bibitem{AdVirgo}
F~Acernese et~al., Classical and Quantum Gravity, {\bf 32}(2), 024001 (dec 2014).

\bibitem{GW170817}
B.~P. Abbott et~al., Phys. Rev. Lett., {\bf 119}, 161101 (Oct 2017).

\bibitem{Multimessenger}
B.~P. Abbott et~al., The Astrophysical Journal Letters, {\bf 848}(2), L12 (oct 2017).

\bibitem{GWTC3-arxiv}
{The LIGO Scientific Collaboration}, {the Virgo Collaboration}, and {the KAGRA Collaboration}, arXiv e-prints, page arXiv:2111.03606 (November 2021).

\bibitem{KAGRA1}
T~Akutsu et~al., Progress of Theoretical and Experimental Physics, {\bf 2018}(1), 013F01 (01 2018),  {{https://academic.oup.com/ptep/article-pdf/2018/1/013F01/23570266/ptx180.pdf}}.

\bibitem{Impact_of_upconverted_scattered_light}
David~J Ottaway, Peter Fritschel, and Samuel~J. Waldman, Opt. Express, {\bf 20}(8), 8329--8336 (Apr 2012).

\bibitem{GW190521}
R.~Abbott et~al., Physical Review Letters, {\bf 125}(10) (September 2020).

\bibitem{PE_for_CB}
J.~Veitch et~al., Physical Review D, {\bf 91}(4) (February 2015).

\bibitem{pyCBC}
Samantha~A. Usman et~al., Class. Quant. Grav., {\bf 33}(21), 215004 (2016),  {{arXiv:1508.02357}}.

\bibitem{pycbc_inference_PE}
C.~M. Biwer and Others, Publications of the Astronomical Society of the Pacific, {\bf 131}(996), 024503 (January 2019).

\bibitem{yamaT}
Takahiro Yamamoto, Kazuhiro Hayama, Shuhei Mano, Yousuke Itoh, and Nobuyuki Kanda, Phys. Rev. D, {\bf 93}, 082005 (Apr 2016).

\bibitem{sensitive_test_non_Gaussianty}
Ronaldas Macas and Andrew Lundgren, Physical Review D, {\bf 108}(6) (sep 2023).

\bibitem{Virgo_detchar}
F~Acernese et~al., Classical and Quantum Gravity, {\bf 40}(18), 185006 (aug 2023).

\bibitem{LIGO_detchar_O3}
D~Davis et~al., Classical and Quantum Gravity, {\bf 38}(13), 135014 (jun 2021).

\bibitem{Yuzurihara:detchar}
Hirotaka Yuzurihara, PoS, {\bf ICRC2023}, 1564 (2023).

\bibitem{PTEP_detchar}
T~Akutsu et~al., Progress of Theoretical and Experimental Physics, {\bf 2021}(5), 05A102 (02 2021).

\bibitem{gwpy}
Duncan~M. Macleod, Joseph~S. Areeda, Scott~B. Coughlin, Thomas~J. Massinger, and Alexander~L. Urban, SoftwareX, {\bf 13}, 100657 (2021).

\bibitem{gwpy_doi}
Duncan Macleod et~al.,
\newblock gwpy/gwpy: Gwpy 3.0.7 (October 2023).

\bibitem{DQSEGDB}
Ryan~P. Fisher et~al., SoftwareX, {\bf 14}, 100677 (2021).

\bibitem{qtransform}
S~Chatterji, L~Blackburn, G~Martin, and E~Katsavounidis, Classical and Quantum Gravity, {\bf 21}(20) (September 2004).

\bibitem{pyCBC_soft}
Alex Nitz et~al.,
\newblock gwastro/pycbc: v2.3.2 release of pycbc (November 2023).

\bibitem{scatter_virgo}
T~Accadia et~al., Classical and Quantum Gravity, {\bf 27}(19), 194011 (sep 2010).

\bibitem{scatter_LIGO}
S~Soni et~al., Classical and Quantum Gravity, {\bf 38}(2), 025016 (jan 2021).

\end{thebibliography}

\appendix

\section{The PSD of Gaussian noise}
\label{gauss_noise}
In Gaussian noise, the real and imaginary parts of the Fourier coefficients $X_k$ independently follow a Gaussian distribution $N(0,\sigma_k^2)$.
$Y_k$ and $Z_k$ are the real and imaginary parts of $X_k$, respectively.
The probability density functions of $Y_k$ and $Z_k$ are
\begin{align}
   f_{Y_k}(y) = \frac{1}{\sqrt{2\pi\sigma_k^2}} \exp\left\{-\frac{y^2}{2\sigma_k^2}\right\}, f_{Z_k}(z) = \frac{1}{\sqrt{2\pi\sigma_k^2}} \exp\left\{-\frac{z^2}{2\sigma_k^2}\right\}.
   \nonumber
\end{align}

The one-sided PSD $P_k$ is defined as follows :
\begin{align}
    P_k = \frac{2 |X_k|^2}{T_\mathrm{FFT}} = \frac{2 (Y_k^2+Z_k^2)^2}{T_\mathrm{FFT}}. \nonumber
\end{align}
For simplicity, we assume $T_\mathrm{FFT} =2$, and rewrite $P_k = Y_k^2+Z_k^2$. 

When $Y_k$ and $Z_k$ independently follow $N(0,\sigma_k^2)$, the cumulative distribution function $F(p_k)$ for $P_k = Y_k^2+Z_k^2$ is  
\begin{align*}
    F(p_k)&= P(P_k \le p_k)\\
    &= P(Y_k^2+Z_k^2 \le p_k)\\
    &= \frac{1}{2\pi \sigma_k^2} \int_{y_k^2+z_k^2\le p_k} \exp\left({-\frac{y_k^2+z_k^2}{2\sigma_k^2}}\right)dy_kdz_k\\
    &= \frac{1}{2 \sigma_k^2} \int_{0}^{p_k} \exp\left({-\frac{u}{2\sigma_k^2}}\right) du.
\end{align*}
The last equal sign is calculated by converting to polar coordinates and transforming the variable $u=r^2$. 
Since the probability density function $f(p_k)$ for $P_k$ is obtained by differentiating  $F(p_k)$, $f(p_k)$ is

\begin{align*}
    f(p_k)&= \frac{d}{dp_k} F(p_k) \\
    &= \frac{1}{2 \sigma_k^2}  \exp\left({-\frac{p_k}{2\sigma_k^2}}\right).
\end{align*}
This is the exponential distribution of parameter $\lambda = 1/2\sigma_k^2$. Therefore, the PSD of Gaussian noise follows an exponential distribution.

\end{document}